# Digital Surveillance Systems for Tracing COVID-19: Privacy and Security Challenges with Recommendations

Molla Rashied Hussein
*Dept. of Comp. Sci. and Engg.*
*University of Asia Pacific*
Dhaka, Bangladesh
mrh.cse@uap-bd.edu

Abdullah Bin Shams
*Dept. of Electrical and Comp. Engg.*
*University of Toronto*
Toronto, Canada
abdullahbinshams@gmail.com

Ehsanul Hoque Apu
*Dept. of Biomedical Engineering*
*Michigan State University*
East Lansing, MI, USA
hoqueapu@msu.edu

Khondaker Abdullah Al Mamun
*Dept. of Comp. Sci. and Engg.*
*United International University*
Dhaka, Bangladesh
mamun@cse.uiu.ac.bd

Mohammad Shahriar Rahman
*Dept. of Comp. Sci. and Engg.*
*University of Liberal Arts Bangladesh*
Dhaka, Bangladesh
shahriar.rahman@ulab.edu.bd

*Abstract*—Coronavirus disease 2019, i.e. COVID-19 has imposed the public health measure of keeping social distancing for preventing mass transmission of COVID-19. For monitoring the social distancing and keeping the trace of transmission, we are obligated to develop various types of digital surveillance systems, which include contact tracing systems and drone-based monitoring systems. Due to the inconvenience of manual labor, traditional contact tracing systems are gradually replaced by the efficient automated contact tracing applications that are developed for smartphones. However, the commencement of automated contact tracing applications introduces the inevitable privacy and security challenges. Nevertheless, unawareness and/or lack of smartphone usage among mass people lead to drone-based monitoring systems. These systems also invite unwelcomed privacy and security challenges. This paper discusses the recently designed and developed digital surveillance system applications with their protocols deployed in several countries around the world. Their privacy and security challenges are discussed as well as analyzed from the viewpoint of privacy acts. Several recommendations are suggested separately for automated contact tracing systems and drone-based monitoring systems, which could further be explored and implemented afterwards to prevent any possible privacy violation and protect an unsuspecting person from any potential cyber attack.

## I. INTRODUCTION

Coronavirus Disease 2019 (COVID-19) has immensely altered the global economy by confining people from daily commercial activities [1]. With no specific vaccine or medicine for preventing or curing [14], currently, the best approach to avoid COVID-19 is not to get ourselves exposed to it. Having close contact with other persons, which is approximately less than six feet [2], can significantly increase the likelihood of inhaling respiratory droplets from the infected persons by coughing, sneezing, or even talking [2]. Social distancing has to be preserved between all, as recent studies [15] have shown that asymptomatic people could silently propagate the disease without raising any suspicion.

This social distancing invokes public health specialists and workers to join with the technical researchers to implement the digital surveillance systems (DSSs) to monitor people digitally, since the process is quite tedious to carry out manually. One of the foremost DSSs is known as automated contact tracing (ACT). To detect individuals who have recently come close in contact with COVID-19 positive persons, ACT systems automates the traditional manual interview with the affected individuals which are administered by the health authorities. Instead of gathering the contacts the affected individual had with other individuals for the last 2-3 weeks through extensive interviews, the ACT systems using smartphone applications (apps) can be an excellent alternative which has been introduced, developed, and even deployed in some countries [3]. Also, it is quite challenging for people to reminisce of their last 2-3 weeks precisely when they can also affect other unknown persons in public places. Moreover, these interviews need a skilled workforce with continuous rigorous training, which is sometimes not feasible.

Along with ACT apps that use smartphones and their location tracing via global positioning system (GPS), wireless-fidelity (Wi-Fi), built-in Bluetooth interface to communicate with and identify proximity with nearby smartphones, also make the privacy and security challenges emerging to a certain level of discomfort. The apps could be repurposed to target their users. collected data could be misused, several security issues such as jamming, storage and power drain attacks, active and passive eavesdroppers, hacking security cameras used in ACT apps are just a few of them.

Another prominent DSSs are drone-based monitoring systems. Drone belongs to a class of unmanned aerial vehicle (UAV) and it is the most common available UAV nowadays. The operational control of a drone can be divided into three categories [16], namely, remote pilot control, i.e. fully controlled manually by an operator; remote supervised control where the drone can perform a defined task autonomously but also allows for a human intervention at the same time, if necessary; and autonomous control in which the drone carries out a defined task on its own.

However, drone surveillance may cause the violation of privacy if the image or video data is downloaded by an intruder, images or videos of an individual get stolen from a drone, Joint Photographic Experts Group (JPEG) image format that contains location and time information [17] falls into the wrong hands. Moreover, hijacking, GPS signal spoofing, control signal hacking, sending too many wireless connection requests can affect drone-based monitoring systems.

This paper is organized as follows: Section I introduces the research topic; after defining the terminologies in section II, the relevant DSS tracing approaches, i.e. ACT apps and protocols are discussed along with drone-based monitoring systems in section III; the emerging privacy and security challenges are discussed and summarized in section IV and V respectively; merits and demerits of these diversified approaches are also analyzed in those sections. Finally, various recommendations are suggested to be explored and implemented as the future works in section VI. The paper concludes with section VII.

## II. DEFINITION OF TERMINOLOGIES

Before discussing DSS tracing approaches, three separate approaches to the system architecture for developing DSS must be briefly discussed. First is the centralized approach, next is the decentralized one, and last but not least is the hybrid one that comprises both the centralized as well as the decentralized architecture features.

In the centralized architecture, the user has to sign oneself up to a central server. The server then automatically creates a privacy-preserving Temporary ID (TempID) for each of the registered devices. TempID is encrypted with a secret key afterwards and known only to the central server's authority. Then it is sent back to the device. Devices exchange these TempIDs through Bluetooth encounter messages after coming in close contact with one another. Whenever a user has been tested COVID-19 positive, all of the stored encounter messages can be voluntarily uploaded to the central server. The server then maps the TempIDs from the messages of individuals to detect contacts that may be at any probable risk.

Contrary to the approach of the centralized architecture, the decentralized architecture only involves servers as a limited medium, more as a bulletin board for the required lookup. It ensures user privacy by creating anonymous identifiers inside the user devices. So, the real user identities are kept secret from both the other users and the server.

The ACT apps can only be successful when the users are assured about their privacy and security while using the app. A naive ACT approach could be building a privacy-agnostic system which is responsible for advertising and exchanging the mobile phone numbers of the users and registers their location time to time by a centralized server. Such a strategy would definitely raise severe privacy and security challenges. The system would most likely be rejected by the common users and thus jeopardize the initial intention.

All ACT apps regardless of having a basis on centralized, decentralized, or hybrid architecture are vulnerable to user misuse. Moreover, ACT apps have failed to detect whether the mobile phone's carrier is someone else or a pet being tied to and running in the park [3].

Prior to discussing privacy and security challenges, various terminologies need to be defined beforehand. First, an adversary is defined as a person who may or may not be tested for COVID-19 and can generate false fear to mass people by spreading fake exposure to COVID-19.

Second, the adversary can attack in several ways. Eavesdropping is the process of gathering data sent by honest users. This can be done by hacking and controlling all mobile phones. Injection is another process by which an adversary is injecting messages by targeting some specific users. This one also involves hacking and controlling mobile phones. Relay attack is an attack where instead of direct attacking, relaying received messages through some other controlled phones to victims' phones. Replay attack or attacking by replaying previously received messages to victims' phones is carried out to take advantage of victims' un-mindfulness [8].

Third, let us take a quick glance towards the privacy acts and laws. The health insurance portability and accountability act (HIPAA) was signed into a privacy law back on August 21, 1996, by then-president Clinton of the United States. For over the last twenty years, the federal department of health and human services (HHS) has released various rule sets within HIPAA's scope and health information technology for economic and clinical (HITECH) act which is under the American recovery and reinvestment act (ARRA), signed into a law by then-president Obama on February 17, 2009 [12].

On the other hand, the European Commission (EC) originated the idea for protecting its citizens on January 25, 2012 regarding the personal data processing and the free movement of such kinds of data all over the internet. The European Union's (EU's) final General Data Protection Regulation (GDPR) was released on May 4, 2016, and was deployed on May 25, 2018 [12]. More about these acts, their applications and suggested recommendations will be discussed in the upcoming sections.

Fourth, the drone mentioned in this literature communicates with the ground control over Wi-Fi or a radio channel [23]. The Wi-Fi channel is predominantly used for a short distance communication, a few hundreds of meters, whereas the long distance communication, thousands of meters, is done using radio waves. Multiple sensors are installed in a drone to achieve an optimal performance. Usually a drone has an on-board camera which is used to gather visual information in terms of either videos or still images.

Multiple sensors are installed in a drone to achieve an optimal performance:

Motion Sensors: Gyroscopes, Accelerometers, and magnetometers are used for an on flight auto stabilization and provide nine degrees of freedom to the drone for a wide range of movements.

Global Positioning System (GPS): GPS is a satellite-based radio navigation system used by the drone to navigate automatically during a flight and to accurately localize its position.

Barometer: This device is used by the drone to estimate its altitude during a flight.

Collision Avoidance Sensors: Ultrasonic sensors are the most common low cost sensors used to detect and avoid obstacles [26]. A more advanced technique relies on computer vision algorithms [27]. In the latter method, data is obtained from multiple on board cameras and analyzed using various image processing tools.

Camera: Usually a drone has an on-board camera which is used to gather visual information in terms of either videos or still images.

### III. LITERATURE REVIEW

Governments in several countries have already employed DSS in the form of ACT apps or programs which use various devices and data sources. To name a few, ACT apps are being implemented in (alphabetically) Bahrain, China, Colombia, Czech Republic, Ghana, India, Israel, Japan, North Macedonia, Norway, Singapore, South Korea and several states in the United States [6]. However, Singapore and Taiwan were prepared for COVID-19 as the learning from the 2003 Severe Acute Respiratory Syndrome (SARS) outbreak was still a fresh memory to recall. Moreover, South Korea planned and organized well due to the lesson learned from the 2015 Middle Eastern Respiratory Syndrome (MERS). Japan, on the other hand, initially utilized the group mentality to inspire over self concern to deploy social distancing successfully [11].

While China imposed their citizens on using the ACT app as compulsory, Singapore relaxed the usage as an elective one. Hong Kong forced to put on electronic wristbands to their COVID-19 positive persons for tracing the movement as well as monitoring whether they are going to leave or already leaving their homes. South Korea sent their regional habitants the government text messages which had included the details of rising cases of COVID-19 pandemic through a central database having anonymous information. Taiwan, along with Israel, utilized their cell tower data. However, while Taiwan used it to employ quarantine geo-fencing, Israel used it for the ACT systems and processes [6].

Now, we focus on some related works focusing on the ACT apps and the associated protocols. First, the TraceTogether app is considered as the pioneer of the implemented ACT apps for mass people. It was endorsed by the Singaporean government back in March 2020. Second, CovidSafe was deployed by the Australian government in the last week of April 2020. CovidSafe abides by the Bluetrace protocol, same as TraceTogether. Third, ROBust and privacy-presERving proximity Tracing protocol, popularly known by its acronym ROBERT, is an ACT app protocol for the centralized approach. It is co-developed by the researchers from Institut National de Recherche en Informatique et en Automatique (INRIA) of France and Fraunhofer of Germany. The StopCovid app, abiding by the ROBERT protocol, has been released by the French government in early June 2020 and lacks enough privacy protection for its users [9].

Aarogya Setu is another ACT app which was implemented in India. It relies on the approach of centralized architecture. However, Apple and Google have used the decentralized architecture approach to co-build a privacy-preserving contact tracing (PPCT) [3]. The authors [7] proposed a PPCT app for COVID-19 using a zero-knowledge protocol for assuring user privacy protection. No user is able to send any fake message to the system for raiding with a false positive attack. The ContactChaser [8] app is designed as such that a health authority is only required to issue group private keys to the users for only one single time. Frequent updating of keys with authority is not necessary. Therefore, the authority finds out the close contacts of affected people with minimum leakage of information.

BeAware app has been designed and implemented by the Kingdom of Bahrain [13]. It tracks down the COVID-19 positive persons by making them wear electronic bracelets, or e-bracelets. It provides the warning signal to the government whenever it suspects any suspicious activity is going on. It comprises Bluetooth with GPS for tracking the person's movement efficiently. The ministry of health (MOH) in Saudi Arabia deployed an app, namely, "Tatamman" to restrict COVID-19 transmission [18]. The app not only provides friendly services to the detected cases for follow-up work and test results from laboratories, but also to the users who came close in contact with the affected persons [18].

Moving back to drone-based monitoring systems, Draganfly, a Canada-based advanced drone manufacturing company, has developed a "Pandemic Drone" that can scan a crowd and identify people with higher temperatures to indicate fever, monitor social distancing and detect coughing by an individual [19]. By measuring the rise in percentages of detected people with the abovementioned traits over many days, an offset of a pandemic may be identified.

Similarly, a startup company based in India, Indian Robotics Solutions (IRS), has developed "Thermal Corona Combat Drone". This drone can detect elevated body temperature, sanitize an area, carry a portable medical box to carry medicines or COVID-19 testing kits and uses a loudspeaker to provide instructions [20].

Moormann proposed an emergency ambulance to deliver medical equipment and medicine supplies for areas hard to reach [21]. Back in 2005, drones were deployed to assess the damages and to check on the state of the survivors post hurricane Katrina.

### IV. PRIVACY CHALLENGES

Traditional methods for protecting privacy have shown a significant lack of integrity regarding privacy and security. Eliminating personal identity information (PII) only or using an anonymous ID is not at all sufficient to assure privacy to the users. Another prime issue for assuring data privacy is managing computational data used in the cryptographic method [10].

The adversary can tag his mobile phone with the ACT app to a carrier, namely a trained animal like dog, or a vehicle or even to a drone. This will broadcast false proximity to the mass people. The adversary may target any specific person and hamper privacy as well. This kind of privacy violation leads to waste of expensive diagnosis resources as well as trust in ACT apps which will be devastating [1].

Moreover, the adversary can attempt to undermine the system efficiency by deducing the sensitive privacy of the honest mobile phone users. To be precise, the adversary tries to provide a false report, i.e. compel the health authority into falsely believing that a person is a close contact of any affected person. The adversary can also try to track a person down by following public information and mapping the hacked or controlled mobile phones information [8].

The original HIPAA clearly defined that one of the three classes for which privacy regulations were applicable was health care providers who collect and store health information in electronic form [12]. Since then numerous reformations have been applied on it. Usually, after any major incident, HHS issues a press release. For example, protected health information (PHI) of approximately 3,800 individuals was lost by a workforce member of a medical center as the member's device got lost in an airport. The device was neither encrypted nor password protected. Same occurrence could happen to ACT apps also.

While HIPAA focuses on privacy rules which do not specify the concept of authorization, GDPR proposed by the EU specifies the idea of consent [12]. Privacy rules allow the right to revoke, on the other hand GDPR allows right to withdrawal of consent. GDPR's right to erasure is applicable to ACT apps, i.e. users should have rights to be forgotten. This will prevent misuse of stored private data after any epidemic/pandemic ends. Privacy rules of HIPAA do not provide these rights [12] and lead to potential privacy violations in the future.

On the other hand, drone surveillance may introduce the violation of privacy especially if the data in the form of image or video is downloaded by an alien intruder. Images or videos of an individual stolen from a drone during an upload or from the cloud server where the drones store all its data may be used in any malicious way against the individual. Some image formats, such as JPEG contain information about the location and time of the photo in the image header files [17]. Therefore with every stolen picture extra individual privacy will be violated.

## V. SECURITY CHALLENGES

Along with eavesdropping, injection, relay attack and replay attack, the attacks known as power drain attack or storage drain attack can be executed as well to make the attacked device busy or slowing down by sending a large amount of notification. The device drains its power every time it wakes up to receive a notification and uses up its storage to store those notifications. The power and storage drain attacks may not appear as a severe issue, but in the long run may lead to opting out of the ACT apps and pose a severe threat to the society [1].

Also, some strict companies or conservative individuals [4] may not allow the CT process running on their premises. This will block the pseudonym exchange through jamming the respective channels.

Eavesdropping, especially passive or unnoticed eavesdropping attack works well in decentralized systems, for example the decentralized privacy preserving proximity tracing (DP-3T). As used pseudonyms of affected individuals are published without any measure, users can be denominated by joining their successive pseudonyms with deduction of their day-to-day movements [4].

Bluetooth based ACT systems are risky as the adversary can generate false alarms regarding contacts with an affected person using a powerful antenna The adversary can also disrupt or track the user's contact tracing [5].

In case of drone-based monitoring systems, the operation and control of a drone depends on several wireless sensors, e.g. Wi-Fi/Radio Frequency (RF) module, GPS etc. Therefore, it is vulnerable to be hijacked. By spoofing the GPS signal [22-23], the GPS signal used by the drone for its navigation is overlapped with an artificially generated GPS signal. When the simulated signal strength is higher than the actual GPS signals from the satellite, the drone can be maneuvered successfully by feeding in false GPS coordinates. Through hacking the control signal [22], as the radio signal used for communicating with the drone is generally not encrypted, listening to the RF signals, using a RF receptor, the signal can then be easily decoded to gain partial or full control of the drone and any of the on board systems, sensors etc. In case of wireless connection request [24], if the drone is targeted and sent a continuous flow of wireless connection requests within a short time interval, this overloads the central processing unit of the drone and results in a shutdown of the entire system.

Most of the drones are imported or assembled using components manufactured in a foreign country. So there is always a possibility that the state sponsored manufacturing company may use their devices for surveillance, to spy, and extract sensitive information. Once a drone is compromised than other devices or drones on the same network are also exposed to be hijacked.

## VI. RECOMMENDATIONS

We recommend the following to preserve privacy during ACT systems:
  (a) As per GDPR, the right to be forgotten has to be implemented to prevent app data usage after pandemic ends.
  (b) HIPAA is health specific only, whereas GDPR is applicable in a broad spectrum. Still, we have to implement the need basis acts for privacy to some countries, especially the South Asian ones.
  (c) Public concern surveys need to be taken to raise mass awareness.
  (d) As the app documentation is missing, mass people are in the dark regarding their privacy issues. Therefore, proper documentation is needed.
  (e) The authors [4] of DP-3T proposed the concept of beacon secret sharing to reduce and restrict passive eavesdropping. Instead of advertising the actual whole pseudonyms, only fractions of pseudonyms are broadcasted. The other side gathers a required number of shares to deduce the original pseudonym of the sender. This approach makes it quite hard for a publicly located BLE to gather meaningful pseudonyms from a person who is just passing by.

The following steps could be taken to preserve privacy during drone-based monitoring systems:
  (a) Drones can be used for tracing in case of lack of awareness or inadequate usage of smart phones among the mass people.
  (b) Upload compressed photos. If a photo is compressed then most of the information in the image header file is erased to minimize the size of the file [23]. This will ensure that the information of a person's whereabouts are erased from an image providing more privacy if the image is stolen.
  (c) Add noise intentionally to distort the picture and store it in a secured network. This will ensure full privacy of an individual even if the image is stolen by some alien user. But if only one type of noise is added then by estimating the noise level variance from the noise statistics, most of the original image information can be restored [25]. Therefore it would be wise to superimpose at least two types of noise. If the image information is required, it can be recovered by specific algorithms known only to the legal authorities.
(d) Ask consent form people to allow the drones to take pictures if necessary for contact tracing only.
(e) If an image of an individual is taken, the person can be notified by the authority, via SMS or email, about the date, location and time when the image was taken.
(f) An indicator in the form of a red blinking light may be activated while the drone is recording or taking the image of an individual.
(g) A signboard can be installed indicating a drone surveillance zone. This will make people aware about any drone that might capture an image of them for contact tracing.

Recommendations for security challenges regarding ACT systems:
(a) To prevent cyber attack, raising awareness is important. Unsuspecting person is vulnerable to this attack more often; therefore, proper learning should be arranged.
(b) App manual should cover the topic of how to be secured from hackers.
(c) Advertisements or commercials could be prepared to let users know about fraudulent activities.
(d) User misuse can create more security threats than adversarial activities, so users have to be responsible while dealing with sensible data.
(e) As the decentralized architecture is more susceptible towards security threats [4], the centralized or hybrid approaches could be followed.

The following recommendations could be explored for security challenges regarding drone-base monitoring systems:
(a) Wireless charging stations must be provided at specific locations. In this way, the drones can operate for a longer time and need not be flown back to the ground control for recharging the battery.
(b) Every command given to the drone must be encrypted and once the command is received by the drone it should have the algorithm to decode and extract the information. This will ensure that when a drone is hacked it would not respond to commands from other alien sources.
(c) Before a data, image or a video is uploaded to a designated authorized network, a security handshake must be successfully executed to prevent data theft.
(d) The drones can also have a separate emergency channel. In case any foreign user takes control over a drone, a signal may be sent over the emergency channel which will force the drone to stop communicating over the normal channel and will follow a protocol to fly back to the ground control.
(e) Any firmware or software must be updated regularly to minimize the chances of being compromised and any password to access a drone must be changed regularly.

## VII. CONCLUSION

To conclude, it is safe to say that DSSs should be used properly aligned to privacy acts and security measures. Both ACT and drone surveillance could be vulnerable to privacy and security challenges. However, this paper not only summarizes the challenges, but also provides several recommendations to consider. In the near future, these recommendations could be explored further and implemented to enhance the current situation. Since COVID-19 is staying for long and recurring, there is no other option than staying safe. Appropriate usage of DSSs can ensure this safety and constant improvements should be carried out to make smooth operations.